\documentclass[symmetry,article,submit,moreauthors,pdftex]{Definitions/mdpi} 

\firstpage{1} 
\pubvolume{1}
\issuenum{1}
\articlenumber{0}
\pubyear{2021}
\copyrightyear{2020}
\datereceived{} 
\dateaccepted{} 
\datepublished{} 
\hreflink{https://doi.org/} 

\usepackage{lipsum}
\usepackage[final]{changes}
\Title{New Light $H^\pm$ Discovery Channels at the LHC}

\TitleCitation{New light $H^\pm$ discovery channels at the LHC}

\Author{Abdesslam Arhrib $^{1}$, Rachid Benbrik $^{2}$, Mohamed Krab $^{3,}$*, Bouzid Manaut $^{3}$, Stefano Moretti $^{4}$, Yan Wang $^{5}$ and Qi-Shu Yan $^{6,7}$}

\AuthorNames{Abdesslam Arhrib, Rachid Benbrik, Mohamed Krab, Bouzid Manaut, Stefano Moretti, Yan Wang and Qi-Shu Yan}

\AuthorCitation{Arhrib, A.; Benbrik, R.; Krab, M.; Manaut, B.; Moretti, S.; Wang, Y.; Yan, Q.-S.}

\address{%
$^{1}$ \quad Abdelmalek Essaadi University, Faculty of Sciences and techniques, Tanger, Morocco; aarhrib@gmail.com\\
$^{2}$ \quad Laboratoire de Physique Fondamentale et Appliquée Safi, Faculté Polydisciplinaire de Safi, Sidi Bouzid, B.P. 4162, Safi, Morocco.; r.benbrik@uca.ac.ma\\
$^{3}$ \quad Polydisciplinary Faculty, 
Research Team in Theoretical Physics and Materials (RTTPM), Sultan Moulay Slimane University, Beni Mellal 23000, Morocco; b.manaut@usms.ma\\
$^{4}$ \quad School of Physics and Astronomy, University of Southampton, Southampton, SO17 1BJ, United Kingdom; s.moretti@soton.ac.uk\\
$^{5}$ \quad College of Physics and Electronic Information, Inner Mongolia Normal University, Hohhot 010022, PR China; wangyan@imnu.edu.cn\\
$^{6}$ \quad Center for Future High Energy Physics, Chinese Academy of Sciences, Beijing 100049, PR China; \\
$^{7}$ \quad School of Physics Sciences, University of Chinese Academy of Sciences, Beijing 100039, PR China; yanqishu@ucas.ac.cn}

\corres{Correspondence: mohamed.krab@usms.ac.ma}

\abstract{A light charged Higgs boson has been searched for at the Large Hadron Collider (LHC) via top (anti)quark decay, i.e., $t \to b H^+$, if kinematically allowed. In this contribution, we propose new channels for light charged Higgs boson searches via the pair productions $pp\to H^\pm h/A$ and $pp\to H^+ H^-$ at the LHC in the context of the Two-Higgs Doublet Model (2HDM) Type-I. By focusing on a case where the heavy H state is the Standard Model \replaced{(SM)-like one already observed, we investigate}{(SM)---like one already observed---we investigate} the production of the aforementioned charged Higgs bosons and their bosonic decay channels, namely, $H^\pm \to W^\pm h$ and/or $H^\pm \to W^\pm A$. We demonstrate that such production and decay channels can yield substantial alternative discovery channels for $H^\pm$ bosons at the LHC. Finally, we propose eight benchmark points (BPs) to motivate the search for such signatures.}


\begin{document}

\section{Introduction}

With the discovery of a 125 GeV Higgs boson at the Large Hadron Collider LHC \cite{ATLAS:2012yve, CMS:2012qbp} in 2012, the verification of the Standard Model (SM) of particle physics was completed. However, despite its  agreement with the experiment, the SM is certainly not an ultimate theory. Thus, any extension of the SM is  well motivated. One of the simplest \added{and most straightforward} extensions of the SM, which deserves particular attention, is the Two-Higgs Doublet Model (2HDM). The model contains two Higgs doublet fields that can generate masses for all (massive) fermions and gauge bosons. \replaced{The scalar sector of 2HDM contains}{The particle spectrum of the 2HDM consists of} two Charge-Parity (CP)-even Higgs bosons, $h$ and $H$ (conventionally the mass of $h$ is less than the mass of $H$, $M_h < M_H$), one CP-odd Higgs boson, $A$, and a pair of charged Higgs bosons, $H^\pm$ (in addition to the fermions and gauge bosons of the SM).

At the LHC, a light $H^\pm$ boson has been searched for via the decay of a top (anti)quark ($t\bar{t}$) if kinematically allowed. Typically, this process can be calculated using the usual method of factorizing the production process of proton--proton collisions, $pp \to t \bar{t}$, times the decay one, $\bar{t} \to \bar{b} H^-$, in the Narrow-Width Approximation (NWA). However, if the mass of the charged Higgs boson approaches the maximum, this approximation becomes invalid, and thus it is quite appropriate to target the process $pp \to t\bar{b}H^-$ to search instead~\cite{Guchait:2001pi}. This contribution revisits these two $H^\pm$ production channels for the upcoming LHC Run 3 and compares them to the pair productions $pp \to H^\pm h/A$ and $pp \to H^+ H^-$ in the 2HDM Type-I. We show that signatures from such pair productions followed by \replaced{$H^\pm \to W^\pm h$ and/or $H^\pm \to W^\pm A$ decays}{BR(BR($H^\pm \to W^\pm h$) and/or BR($H^\pm \to W^\pm A$)} may lead to new discovery channels for light charged Higgs bosons searches at the LHC.

The contribution is organised as follows. \replaced{First, we briefly describe the 2HDM and its Yukawa scenarios in Section 2. In Section 3 we explain the scan of the parameter space and the applied constraints. We discuss the numerical results and the selected Benchmark Points (BPs) in both Sections 4 and 5, and we finally conclude in Section 6.}{In section 2, we give a brief description of the 2HDM and its Yukawa scenarios. In Section 3, we describe our parameter space scans. The numerical results and a set of Benchmark Points (BPs) are given in Sections 4 and 5, respectively. Finaly, we present some conclusions.}
 
\section{The 2HDM}
The $\mathcal{CP}$-conserving 2HDM scalar potential, which is renormalisable and invariant under $SU(2)_L \otimes U(1)_Y$ with a softly broken $Z_2$ symmetry, can be written as
\begin{eqnarray}
V(\phi_1,\phi_2) &=& m_{11}^2(\phi_1^\dagger\phi_1) +
m_{22}^2(\phi_2^\dagger\phi_2) -
[ m_{12}^2(\Phi_1^\dagger\phi_2)+\text{h.c.}] ~\nonumber\\&& 
+\,\frac{1}{2}\lambda_1(\phi_1^\dagger\phi_1)^2 + \frac{1}{2}\lambda_2(\phi_2^\dagger\phi_2)^2 + 
\lambda_3(\phi_1^\dagger\phi_1)(\phi_2^\dagger\phi_2) ~\nonumber\\&&
+\,\lambda_4(\phi_1^\dagger\phi_2)(\phi_2^\dagger\phi_1) + \frac{1}{2}\lambda_5[(\phi_1^\dagger\phi_2)^2+\text{h.c.}],
\end{eqnarray}
where $m_{11}^2$, $m_{22}^2$ and $m_{12}^2$ are squared mass parameters, and $\lambda_{1-5}$ are dimensionless coupling parameters. $\phi_{1,2}$ are the Higgs doublet fields with $v_{1,2}$ their respective Vacuum Expectation Values (VEVs) such that $v_1^2 +v_2^2 = v^2 \simeq (246~\,\text{GeV})^2$ (where $v$ is the SM Higgs VEV). \replaced{Using the two minimization conditions of the potential, $m_{11}^2$, $m_{22}^2$ and $\lambda_{1-5}$ can be substituted by $v_{1,2}$, the physical mass eigenstates and $\sin(\beta-\alpha)$, where $\alpha$ and $\beta$ are the mixing angles.}{By using the two minimisation conditions of the potential, $m_{11}^2$ and $m_{22}^2$ as well as the quartic couplings $\lambda_{1-5}$ can be substituted by $v_1$ and $v_2$, and the four physical Higgs boson masses and mixing parameter of the neutral states $\sin(\beta-\alpha)$, where $\beta$ is defined by $\tan\beta = v_2/v_1$ while $\alpha$ is the mixing angle between the two neutral Higgs bosons $h$ and $H$, respectively.} Thus, we are left with only 7 independent parameters:

\begin{eqnarray}
M_h, \, M_H, \, M_A, \, M_{H^\pm}, \,\alpha, \,\tan\beta \,\,\text{and}\,\, m_{12}^2.
\end{eqnarray}

In the Yukawa sector, though, the Flavor Changing Neutral Currents (FCNCs) can be induced at the tree level if both the Higgs doublets of the general 2HDM couple to all fermions. To avoid FCNCs, which would be inconsistent with the experiment, a $Z_2$ symmetry can be enforced in such a way that each fermion type ($u,d,l$) acquires mass from one of the Higgs doublets. Thus, there are four possible Types of 2HDM \cite{Branco:2011iw}. In the 2HDM Type-I, the fermions acquire mass via the interaction with the doublet $\phi_2$ as in the SM. In the 2HDM Type-X (or lepton-specific), the charged leptons acquire mass from $\phi_1$ while all quarks receive mass from $\phi_2$. In the 2HDM Type-II, up-type quarks acquire mass through their interaction with $\phi_2$ while down-type quarks and charged leptons acquire mass through their interaction with $\phi_1$. Finally, in the 2HDM Type-Y (or flipped), the up-type quarks and charged leptons receive mass from $\phi_2$ while down-type quarks receive mass from $\phi_1$.
Here, though, \replaced{we will only consider the 2HDM Type-I.}{we are interested only in 2HDM Type-I.}


The Yukawa Lagrangian which describes \replaced{the coupling of the neutral and charged Higgs bosons to quarks and leptons}{the neutral and charged Higgs couplings to fermions} can be written as \cite{Branco:2011iw}:
\begin{eqnarray}
- {\mathcal{L}}_{\rm Yukawa} = \sum_{f=u,d,l} \left(\frac{m_f}{v} \kappa_f^h \bar{f} f h + 
\frac{m_f}{v}\kappa_f^H \bar{f} f H 
- i \frac{m_f}{v} \kappa_f^A \bar{f} \gamma_5 f A \right) + \nonumber \\
\left(\frac{V_{ud}}{\sqrt{2} v} \bar{u} (m_u \kappa_u^A P_L +
m_d \kappa_d^A P_R) d H^+ + \frac{ m_l \kappa_l^A}{\sqrt{2} v} \bar{\nu}_L l_R H^+ + H.c. \right),
\label{Yukawa-1}
\end{eqnarray}
where $m_f$ ($f=u,d,l$) are the masses of the fermions and $\kappa_f^S$ are the Yukawa couplings, \replaced{which are given in Table 1 for Type-I. $V_{ud}$ denotes the}{which are listed in Table \ref{yuk_coupl} for the 2HDM Type-I of interest here. $V_{ud}$ represents a} Cabibbo-Kobayashi-Maskawa (CKM) matrix element, and $m_u$ and $m_d$ are the masses of up and down quarks, respectively. $P_{L,R}$ \replaced{represent the left- and right-handed}{denote the left- and right-handed} projection operators.
\begin{table}[!t]
		\begin{center}
		\begin{tabular}{||c|c|c|c||} \hline
			  & $\kappa_u^{S}$ &  $\kappa_d^{S}$ &  $\kappa_\ell^{S}$  \\   \hline
			$h$~ 
			& ~ $  \cos\alpha/ \sin\beta$~
			& ~ $  \cos\alpha/ \sin\beta$~
			& ~ $  \cos\alpha/ \sin\beta $~ \\
			$H$~
			& ~ $  \sin\alpha/ \sin\beta$~
			& ~ $  \sin\alpha/ \sin\beta$~
			& ~ $  \sin\alpha/ \sin\beta$~ \\
			$A$~  
			& ~ $  \cot \beta $~  
			& ~ $  -\cot \beta $~  
			& ~ $  -\cot \beta $~  \\ \hline
		\end{tabular}
	\end{center}
	\caption{Yukawa couplings \replaced{of the neutral Higgs bosons $h$, $H$ and $A$ to quarks and leptons in 2HDM Type-I.}{of the fermions ($u,d,l$) to the neutral Higgs bosons ($h, H, A$) in the 2HDM Type-I. Here $c_x$ and $s_x$ ($x=\alpha, \beta$) refer to $\cos x$ and $\sin x$, respectively.}}
	\label{yuk_coupl}	
\end{table}	

\section{Parameter Space Scans}
\replaced{In what follows, we perform a broad scan of the following 2HDM Type-I parameter space, where the H state is assumed to be the observed SM-like Higgs at the LHC in $2012$ with mass fixed to $125$ GeV,}{As previously mentioned, we assume that the observed SM-like Higgs particle at CERN in 2012 is the $H$ state, for which $m_H = 125$ GeV. We then perform a broad scan over the following 2HDM Type-I parameter space:}
\begin{eqnarray}
&&M_h = 10 ~\text{--}~ 120~\text{GeV}; ~~M_H = 125~\text{GeV}; ~~M_A = 10~\text{--}~120~\text{GeV}; ~~M_{H^\pm} = 80~\text{--}~170~\text{GeV} \nonumber\\ 
&& \tan\beta = 2~\text{--}~60; ~~\sin(\beta-\alpha) = (-0.3)~\text{--}~(-0.05); ~~m_{12}^2 = 0~\text{--}~M_{H}^2\sin\beta\cos\beta. 
\end{eqnarray}
In the scan, \replaced{the theoretical and experimental constraints  are taken into account. \texttt{2HDMC}  \cite{Eriksson:2009ws} is used to check unitarity, perturbativity, vacuum stability and the electroweak oblique parameters ($S$, $T$ and $U$). \texttt{HiggsBounds-5.9.0} \cite{Bechtle:2020pkv} and \texttt{HiggsSignals-2.6.0} \cite{Bechtle:2020uwn} are both used to enforce the exclusion bounds at 95\% Confidence Level (CL) from Higgs boson searches at LEP, Tevatron and LHC, and to check agreement with SM-like Higgs boson measurements, respectively. Constraints from flavour physics are tested using the public code \texttt{SuperIso v4.1} \cite{Mahmoudi:2008tp}.}{all up-to-date theoretical and experimental constraints are taken into account thanks to \texttt{2HDMC}  \cite{Eriksson:2009ws}, that has been used to check unitarity, perturbativity, vacuum stability as well as oblique parameters $S$, $T$ and $U$, as well as \texttt{HiggsBounds-5.9.0} \cite{Bechtle:2020pkv} and \texttt{HiggsSignals-2.6.0} \cite{Bechtle:2020uwn}, that have been used to enforce the exclusion bounds at 95\% Confidence Level (CL) from Higgs boson searches at colliders (LEP, Tevatron and LHC) and compliance with SM-like Higgs boson measurements, respectively. Constraints from B-physics observables are checked using the code \texttt{SuperIso v4.1} \cite{Mahmoudi:2008tp}.}

\section{Results}

In the present contribution, we target the signatures\footnote{We are only interested here in the $4b$ final states. See Ref. \cite{Arhrib:2021xmc} for the $2b2\tau$ and $4\tau$ final states.} of light charged Higgs bosons from processes involving top quarks and di-Higgs processes, i.e., $gg, q\bar{q} \to t\bar{t}\to t\bar{b}H^-$ + c.c. (NWA), $gg, q\bar{q} \to t\bar{b}H^-$ + c.c., $q\bar{q} \to H^+ H^-$ plus $q\bar{q}' \to H^+h/A$ + c.c. \replaced{taking into account their either $W^\pm h$ or $W^\pm A$ decays,}{and investigate their bosonic decays, such as $H^\pm \to W^\pm h$ and/or $H^\pm \to W^\pm A$,} where the $h$ and $A$ decay into a pair of bottom quarks. Relevant LHC signatures are summarised in Table \ref{tab2}. 

\begin{table}[h]
    \begin{center}
\begin{tabular}{||l|l||} \hline
	& Higgs production and decay process\\ \hline
	$\sigma^{h_i}_{2t}(2W + 4b)$  &  2 $\sigma_{t\bar{t}} \times {\rm BR}(t \to b H^+)\times {\rm BR}(\bar{t} \to \bar{b} W^-)\times {\rm BR}(H^\pm \to W^\pm h_i) \times {\rm BR}(h_i \to b\bar{b})$ \\ 
	\hline 
	$\sigma^{h_i}_{t}(2 W + 4b)$  & $ \sigma(pp \to t\bar{b}H^-)\times {\rm BR}(t \to b W^+)\times {\rm BR}(H^\pm \to W^\pm h_i) \times {\rm BR}(h_i \to b\bar{b})$ \\
	\hline 
	$\sigma^{h_i}_{h_i}(2W+4b)$ & $\sigma(pp \to H^+ H^-)\times {\rm BR}(H^\pm \to W^\pm h_i)^2 \times  {\rm BR}(h_i \to b\bar{b})^2$  \\ \hline
	~$\sigma^{h_i}_{h_i}(W+4b)$ &  $\sigma(pp \to H^\pm h_i)\times {\rm BR}(H^\pm \to W^\pm h_i) \times {\rm BR}(h_i \to b\bar{b})^2$  \\
	\hline 
\end{tabular}
	\caption{\replaced{Charged Higgs bosons production modes}{The production processes of charged Higgs bosons} and their final states. $\sigma_{t\bar{t}}$ denotes the production process of proton-proton collisions, $pp \to t\bar{t}$, and BR refers to the branching ratio. Here, $h_i$ ($i=1, 2$) refers to $h_1 = h$ and $h_2 = A$.} \label{tab2}
	\end{center}
\end{table}
\begin{figure}[!t]
	\centering
	\includegraphics[scale=0.6]{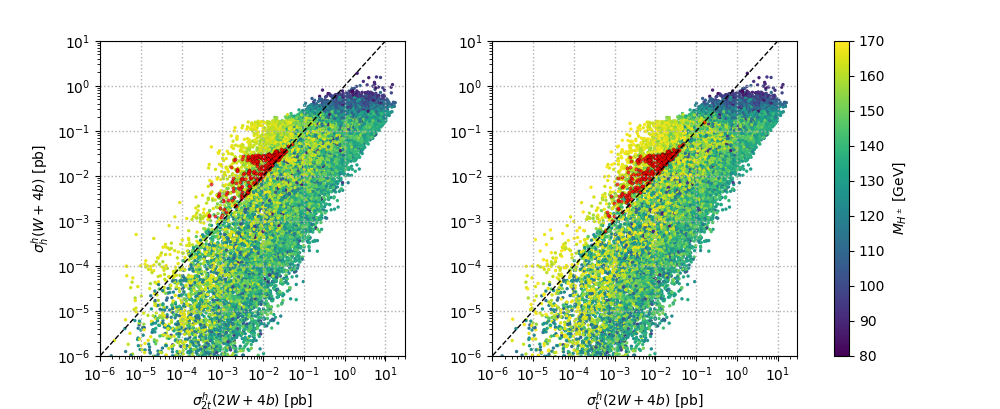}
	\caption{\replaced{Total cross section}{Values of}  $\sigma(pp \to H^\pm h)\times {\rm BR}(H^\pm \to W^\pm h)\times {\rm BR}(h \to b\bar{b})^2$ \replaced{are showed against those}{are compared with those of} 
		2 $\sigma_{t\bar{t}} \times {\rm BR}(t \to b H^+)\times {\rm BR}(\bar{t} \to \bar{b} W^-)\times {\rm BR}(H^\pm \to W^\pm h)\; \times $ 
		${\rm BR}(h \to b\bar{b})$ 
		(left) and 
		$ \sigma(pp \to t\bar{b}H^-)\times {\rm BR}(t \to b W^+)\times {\rm BR}(H^\pm \to W^\pm h)\; \times $ 
		$ {\rm BR}(h \to b\bar{b})$ 
		(right). \replaced{The red points refer to the total cross section}{The red points identify the values of} $\sigma(pp \to H^+ H^-)\times {\rm BR}(H^\pm \to W^\pm h)^2 \times {\rm BR}(h \to b\bar b)^2$
		\replaced{which are also large compared to the two top production processes.}{which also exceeds those of the latter two top mediated processes, respectively.} 
		\replaced{With the mass of charged Higgs boson, $M_{H^\pm}$, indicated by the colour map.}{The colour bar denotes the mass of the charged Higgs boson.}} \label{fig1}
\end{figure}
\begin{figure}[!h]
	\centering
	\includegraphics[scale=0.6]{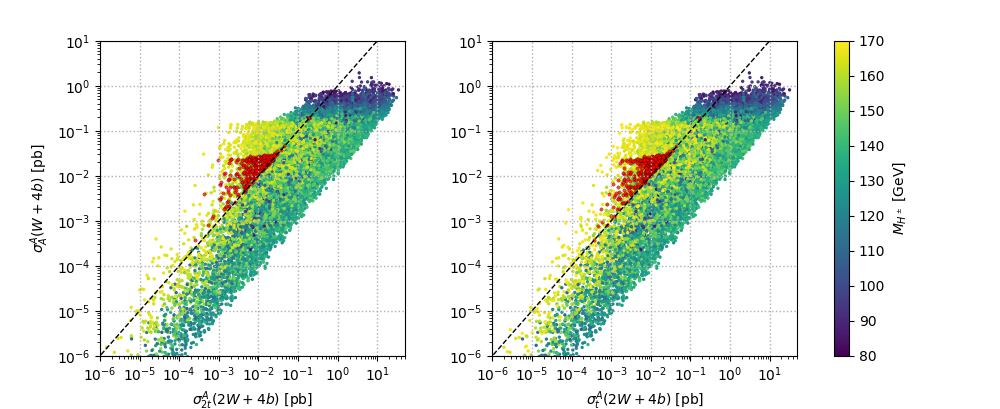}
	\caption{\replaced{Total cross section}{Values of}  $\sigma(pp \to H^\pm A)\times {\rm BR}(H^\pm \to W^\pm A)\times {\rm BR}(A \to b\bar{b})^2$ \replaced{are showed against those}{are compared with those of} 
		2 $\sigma_{t\bar{t}} \times {\rm BR}(t \to b H^+)\times {\rm BR}(\bar{t} \to \bar{b} W^-)\times {\rm BR}(H^\pm \to W^\pm A)\; \times $ 
		${\rm BR}(A \to b\bar{b})$ 
		(left) and 
		$ \sigma(pp \to t\bar{b}H^-)\times {\rm BR}(t \to b W^+)\times {\rm BR}(H^\pm \to W^\pm A)\; \times $ 
		$ {\rm BR}(A \to b\bar{b})$ 
		(right). \replaced{The red points refer to the total cross section}{The red points identify the values of} $\sigma(pp \to H^+ H^-)\times {\rm BR}(H^\pm \to W^\pm A)^2 \times {\rm BR}(A \to b\bar b)^2$
		\replaced{which are also large compared to the two top production processes.}{which also exceeds those of the latter two top mediated processes, respectively.} 
		\replaced{With the mass of charged Higgs boson, $M_{H^\pm}$, indicated by the colour map.}{The colour bar denotes the mass of the charged Higgs boson.}} \label{fig2}
\end{figure}
In what follows, we show the production rates of relevant final states from different scenarios. In Figure \ref{fig1}, we compare \replaced{$W+4b$}{$\sigma(W+4b)$} and \replaced{$2W+4b$}{$\sigma(2W+4b)$} signatures from $pp \to H^\pm h \to W^\pm hh$ and $pp \to H^+ H^- \to W^+ W^- hh$ with $\sigma_{2t}(2W+4b)$ \added{(left panel)} and $\sigma_{t}(2W+4b)$ (right panel) ones from the two top (anti)quark processes. Analogously to Figure \ref{fig1}, the same signatures from $pp \to H^\pm A \to W^\pm A A$ and \replaced{$p p \to H^+ H^- \to W^+ W^- A A$}{$p p \to H+ H- \to W+ W- A A$} are compared with those from processes involving the top (anti)quark in Figure \ref{fig2}. From these plots, it is therefore clear that signatures from di-Higgs processes can yield substantial alternative discovery modes for charged Higgs bosons at the LHC in the context of the 2HDM Type I.
\section{Benchmark Points}
In order to encourage future searches for light charged Higgs boson via such new channels, we propose 8 BPs for the 2HDM Type-I. Such BPs are presented in Table \ref{BPsTypeI}. In our selected BPs, notice that we take also into account the case where the mass of the charged Higgs is larger than the top one. \replaced{The total cross section of the final states $2W+4b$ and $W+4b$ from both di-Higgs and the two top (anti)quarks are given herein.}{The cross sections of the $2W + 4b$ and $W+4b$ signatures from both di-Higgs and top (anti)quark processes are also given herein.}

In BP1, for instance, the cross-section rate of the $2W+4b$ signature from the top (anti)quark processes can only reach $5$ fb\footnote{we refer here to the $pp \to t\bar b H^-$ + c.c. rates}, whereas the cross-section rate of the $2W+4b$ signature from the pair production of $H^\pm$ bosons is $\approx$23.1 fb. Moreover, the cross-section rate \replaced{$\sigma{(W+4b)}$}{sigma(W + 4b)} from the $h/A$-associated $H^\pm$ production can reach values of around $174$ fb, which are much larger than the rates of $\sigma(2W+4b)$ from charged Higgs pair production. This behavior is well illustrated in Figure \ref{rates}. For other BPs, the cross-section rates of the \replaced{$2W+4b$ and $W+4b$ signatures}{$(2)W+4b$} from different production processes are also shown in Figure \ref{rates}.

\begin{table}[hbtp]
	\begin{center}
	\begin{tabular}{||l|l|l|l|l|l|l|l|l||}\hline
		~\, Parameters & ~\,BP1 & \,~BP2   & \,~BP3 & \,~BP4 & \,~BP5 & \,~BP6 & \,~BP7  & \,~BP8\\\hline
		~\qquad $M_h$ &  ~$91.00$ & ~$96.84$ & $103.34$ & ~$99.61$ & ~$95.57$ & ~$94.00$ & ~$94.00$ & ~$94.00$\\
		~\qquad $M_H$ &  $125.00$ & $125.00$  & $125.00$  & $125.00$ & $125.00$  & $125.00$ & $125.00$ & $125.00$\\
		~\qquad $M_A$ &  $102.04$ & $112.35$  & ~$93.80$  & ~$88.98$ & ~$94.41$  & $105.00$ & $105.00$ & $105.00$\\  
		~\qquad $M_{H^\pm}$ &  $167.02$ & $166.34$  & $161.02$  & $169.46$ & $167.02$ & $176.00$ & $186.00$ & $196.00$\\ 
		\quad $\sin(\beta-\alpha)$ & $-0.18$ & $-0.11$ & $-0.19$  & $-0.06$ & $-0.09$ & $-0.09$ & $-0.09$ & $-0.09$\\ 
		\qquad $\tan\beta$ & ~$40.87$ & ~$58.17$ & ~$54.79$ & ~$39.10$ & ~$32.44$ & ~$30.00$ & ~$30.00$ & ~$30.00$\\
		~\qquad $m_{12}^2$ & $204.22$ & $161.85$ & $196.73$  & $252.94$ & $277.81$ & $294.00$ & $294.00$ & $294.00$ \\ \hline
		
		~ $\sigma^{h}_{2t}(2W + 4b)$ & \,~$2.30$ & \,~$1.65$ & \,~$2.06$ & \,\,~~$-$ & \,~$2.42$ & \,\,~~$-$ & \,\,~~$-$ & \,\,~~$-$\\
		~ $\sigma^{h}_{t}(2W+4b)$ & \,~$3.85$ & \,~$2.35$ & \,~$2.26$ & \,~$0.85$ & \,~$3.84$ & \,~$5.03$ & \,~$4.68$ & \,~$3.52$\\ 
		~ $\sigma^{A}_{2t}(2 W + 4b)$ & \,~$0.70$ & \,~$0.25$ & \,~$4.63$ & \,\,~~$-$ &  \,~$2.47$ & \,\,~~$-$ & \,\,~~$-$ & \,\,~~$-$\\
		~ $\sigma^{A}_{t}(2W+4b)$ & \,~$1.17$ & \,~$0.36$ & \,~$5.07$ & \,~$3.03$ &  \,~$3.92$ & \,~$0.83$ & \,~$0.44$ & \,~$1.08$\\ \hline 
		
		~~$\sigma^{h}_{h}(2W+4b)$ & \,$13.58$ & \,$15.99$ & \,~$2.29$ & \,~$0.97$ & \,~$5.38$ & \,$14.08$ & \,$13.27$ & \,~$7.35$\\	
		~~$\sigma^{h}_{A}(2W+4b)$ & \,~$4.13$ & \,~$2.44$ & \,~$5.14$ & \,~$3.46$ & \,~$5.50$ & \,~$2.32$ & \,~$1.25$ & \,~$2.24$\\
		~~$\sigma^{A}_{A}(2W+4b)$ & \,~$1.26$ & \,~$0.37$ & \,$11.55$ & \,$12.35$ & \,~$5.62$ & \,~$0.38$ & \,~$0.12$ & \,~$0.68$\\ 
		~~$\sigma^{A}_{h}(2W+4b)$ & \,~$4.13$ & \,~$2.44$ & \,~$5.14$ & \,~$3.46$ & \,~$5.50$ & \,~$2.32$ & \,~$1.25$ & \,~$2.24$\\ \hline
		\,~~$\sigma^{h}_{h}(W+4b)$ & \,$75.88$ & \,$77.61$ & \,$26.47$  & \,$17.68$ & \,$46.00$ & \,$73.25$ & \,$68.00$ & \,$48.81$\\
		\,~~$\sigma^{h}_{A}(W+4b)$ & \,$23.07$ & \,$11.86$ & \,$59.44$  & \,$63.04$ & \,$47.00$ & \,$12.07$ & \,~$6.42$ & \,$14.90$\\		
		\,~~$\sigma^{A}_{A}(W+4b)$ & \,$17.48$ & \,~$6.12$ & \,$64.39$ & \,$69.22$ & \,$43.51$ & \,~$9.16$ & \,~$4.91$ & \,$11.45$\\ 
		\,~~$\sigma^{A}_{h}(W+4b)$ & \,$57.51$ & \,$40.06$ & \,$28.68$ & \,$19.41$ & \,$42.59$ & \,$55.59$ & \,$52.02$ & \,$37.51$\\ \hline	
	\end{tabular}
	\caption{Mass spectra (in GeV), mixing angles and cross sections (in fb) for the selected BPs. (Notice that all these parameters have been discussed above.)} \label{BPsTypeI}
	\end{center}
\end{table}
\end{paracol}

\begin{paracol}{2}
\linenumbers
\switchcolumn

\section{Conclusions}
In this contribution, we have investigated the production  of charged Higgs bosons through $pp \to H^\pm h/A$ and $pp \to H^+ H^-$ at the LHC with $\sqrt{s} = 14$ TeV in the 2HDM Type-I, \replaced{after satisfying all theoretical and experimental constraints.}{after taking into account all up-to-date theoretical and experimental constraints.} By focusing on $H^\pm \to W^\pm h/A$ decays, \replaced{we have suggested the $2W + 4b$ and $W + 4b$ signatures as possible alternative discovery modes.}{we have examined
the final states $2W + 4b$ and $W + 4b$ as potential discovery channels.} We have demonstrated that such signatures could well be the most promising discovery for light $H^\pm$ states. Thus, \replaced{to motivate experimentalists to search for these,}{in order to motivate experimentalists in ATLAS and CMS to search for these,} we have proposed 8 BPs amenable to experimental investigation.

\begin{figure}[h]
	\begin{center}
		\includegraphics[width=0.4\linewidth]{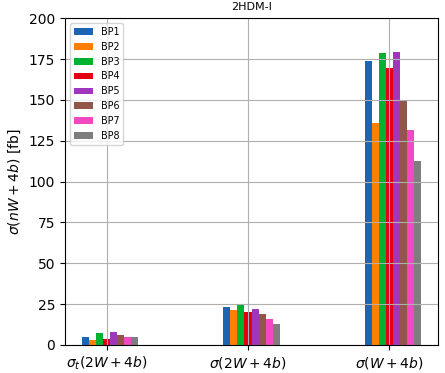}
		\hspace*{3.0truecm}\caption{\replaced{Cross section rates of}{Cross sections of} $2W + 4b$ and $W + 4b$ signatures for the selected BPs.} \label{rates}
	\end{center}
\end{figure}

\vspace{6pt} 

\authorcontributions{All authors have contributed in equal parts to all aspect of this research.}

\funding{The work of A.A., R.B., M.K. and B.M. is supported by the Moroccan Ministry of Higher Education and Scientific Research MESRSFC and
CNRST Project PPR/2015/6. The work of S.M. is supported in part through the NExT Institute and the STFC Consolidated Grant No. ST/L000296/1.
Y. W. is supported by the `Scientific Research Funding Project for
Introduced High-level Talents' of the Inner Mongolia Normal
University Grant No. 2019YJRC001.  Q.-S. Yan's work is supported by the Natural Science Foundation of China Grant No. 11875260.}

\institutionalreview{Not applicable.}

\informedconsent{Not applicable.}

\dataavailability{Not applicable.}

\conflictsofinterest{The authors declare no conflict of interest.} 



\end{paracol}
\reftitle{References}

\end{document}